\ifx\pdfoutput\undefined
	\documentclass[dvips]{cimento}
\else
	\documentclass[pdftex]{cimento}
\fi

\usepackage[dvips]{graphicx}
\usepackage{amssymb}
\usepackage{amsmath}
\usepackage{natbib}
\usepackage[english]{babel}
\bibpunct{[}{]}{;}{a}{}{,}

\DeclareMathOperator{\Diag}{Diag}
\DeclareMathOperator{\eV}{eV}

\title{Effect of neutrino asymmetry on the estimation of cosmological parameters}
\author{M.~Lattanzi\from{inst:icra}}
\instlist{\inst{inst:icra} ICRA - International Center for Relativistic Astrophysics and Physics Department, University of Rome ``La Sapienza'', P.le A. Moro 5, 00185 Rome, Italy}
\PACSes{\PACSit{95.35.+d}{Dark Matter}
\PACSit{98.80.-k }{Cosmology}\PACSit{14.60 -z}{Leptons}}
\begin{document}

\maketitle

\begin{abstract}
The recent analysis of the cosmic microwave background data carried out by the WMAP team seems to show that the sum of the neutrino masses is $<0.7$~eV. However, this result is not model-independent, depending on precise assumptions on the cosmological model. We study how this result is modified when the assumption of perfect lepton symmetry is dropped out.
\end{abstract}

\newcommand{\nue}{\nu_e}
\newcommand{\num}{\nu_\mu}
\newcommand{\nut}{\nu_\tau}

\section{Introduction}
The anisotropy spectrum of the cosmic microwave background is without doubt one of the most important observables quantity in cosmology. Recently, its precise measurement by the WMAP experiment, has lead to a precise determination of the values of the parameters describing our Universe \citep{Spe03}. In particular, the WMAP team finds a very tight bound on the sum of neutrino mass, representing an improvement of roughly one order of magnitude with respect to the laboratory limit. It seems that finally the exciting possibility of performing precision tests in particle physics models using cosmological data, is within our reach. 

Unfortunately, parameter estimation is plagued with the problem of parameter degeneration. With this it is meant that different parameters have similar effect on the microwave background spectrum; the latter cannot be used, then, to distinguish between them. A well known example is the degeneracy between the spectral index $n$ and the reionization redshift $z_{nr}$. Put in another way, we could say that there is not a one-to-one relationship between the parameters and the spectrum. The simplest way to break such a degeneracy is to measure some other quantity; for example, the matter power spectrum.

Another difficulty in dealing with the degeneration of parameters is due to the fact that we do not have an explicit relationship between the parameters and the spectrum. This means that probably there are some degeneracies we are not aware of, and that, even for kwown degeneracies, we do not know exactly how much they influence our results. Thus we are forced to use numerical methods, that are in some cases very time expensive.

Our work deals with the possibility to extract from the microvave background informations on the degree of lepton asymmetry existing in our Universe. In particular, we study how the estimation of cosmological parameters is affected by the existence of such an asymmetry. The paper is structured as follows: in section 2, we review the basis of neutrino physics; in section 3, we briefly recall the basis of the microwave background physics ; in section 4, we describe the techniques for parameter estimation; in section 5 we speak about the role of neutrinos in cosmology; in section 6 we present and discuss our results; finally in section 7 we draw our conclusions.

\section{Neutrino Physics}\label{sec:nuphys}
In this section we review some basic facts of neutrino phyics, with particular regard to the formalism and to the phenomenology of neutrino mixing.
 
Neutrino oscillations are due to the fact that the flavour eigenstates $\nu'\equiv(\nue,\,\num,\,\nut)$, defined on the basis of the interaction with the charged leptons (so that $\nue$ is the partner of the electron, and so on) do not coincide with the mass eigenstates $(\nu\equiv{\nu_1,\,\nu_2,\,\nu_3})$, but are instead related by a unitarity transformation:
\begin{equation}
\nu'=U\nu
\label{eq:mixing}
\end{equation}
where $U$ is the unitary $3\times 3$ mixing matrix, parametrized in terms of three mixing angles $(\theta_{12},\,\theta_{23},\,\theta_{13})$ and one phase $\phi$. Denoting $m_{\mathrm{diag}}\equiv\Diag\,(m_1,\,m_2,\,m_3)$, the effective mass matrix $m_\nu$ is given by:
\begin{equation}
m_\nu=U m_\mathrm{diag}U^T
\end{equation}
It is customary to label the mass eigenstates so that:
\begin{equation}
0<\Delta m^2_{21}<|\Delta m^2_{32}|
\end{equation}
where $\Delta m^2_{ij}=|m|^2_i-|m|^2_j$.
The frequency $\omega_{ij}$ of neutrino oscillations is determined by their mass difference:
\begin{equation}
\omega_{ij}=\frac{|\Delta m^2_{ij}|}{4E}
\end{equation}
The two distinct frequencies measured in the experiments involving atmospheric and solar neutrinos correspond, in terms of the mass eigenstates, to:
\begin{equation}
\Delta m^2_\mathrm{sun}\equiv\Delta m^2_{21},\quad
\Delta m^2_\mathrm{atm}\equiv|\Delta m^2_{32}|
\end{equation}

At the present time, we have the following information about the mass square differences \citep{Al03}:
\begin{eqnarray}
&\Delta m^2_\mathrm{sun}&\sim0.7\cdot10^{-4}\eV\\
&\Delta m^2_\mathrm{atm}&\sim0.2\cdot10^{-2}\eV\nonumber
\end{eqnarray}
and the following about the mixing angles \citep{Al03}:
\begin{eqnarray}
&\theta_{12}&\sim 30^\circ\div35^\circ \\
&\theta_{23}&\sim 45^\circ\nonumber\\
&\theta_{13}&\lesssim 10^\circ\nonumber
\end{eqnarray}
However, we are still missing some important informations: first of all, the absolute values of neutrino masses, and consequently the mass hierarchy. Given the experimental data, there are three possible scenarios:
\begin{alignat}{4}
|m_1|&\sim&\,|m_2|&\sim&\,|m_3|&\gg|m_i-m_j|& \qquad&\textrm{Degenerate}\nonumber \\
|m_1|&\sim&\,|m_2|&\gg&\,|m_3|&&\qquad&\textrm{Inverted hierarchy}\\
|m_3|&\gg&\,|m_1|&\sim&\,|m_2|&&\qquad&\textrm{Normal hierarchy} \nonumber
\end{alignat}

Secondly, we need a more precise determination of the three mixing angles, in particular $\theta_{13}$ for which only an upper bound is known. Finally, it is still unknown if neutrinos are Dirac or Majorana particles. If the latter one is the case, two additional parameters, namely the relative phases $\phi_1$, $\phi_2$ between the Majorana mass terms $m_1$, $m_2$ and $m_3$, are needed.

The tritium $\beta$ decay experiments give the following bound for $m_e$ \citep{We03} \cite{Kr03}:
\begin{equation}
m_e<2.2\eV
\label{eq:m_e bound}
\end{equation}

It should be noted that, since $\nue$ is not a mass eigenstate, its mass can only be defined as an expectation value:
\begin{equation}
m_e^2=\sum_{i=1}^3|U_{e i}|^2m^2_i
\label{eq:m_e}
\end{equation}

Another important, complementary experimental approach is the search for the neutrinoless double $\beta$ decay (in brief, $0\nu\beta\beta$ decay). This process is sensitive to the ``effective'' neutrino mass $m_{ee}$, namely the 11 entry of the mass matrix $m_\nu$:
\begin{equation}
m_{ee}=\left|\sum_{i=1}^3U_{ei}^2\,m_i\right|
\label{eq:m_ee}
\end{equation}
The evidence for $0\nu\beta\beta$ decay, if found, would imply among others that lepton number is violated and that neutrinos are Majorana particles. At the present only an upper bound on $|m_{ee}|$ is known:
\begin{equation}
|m_{ee}|<0.5\eV
\label{eq:m_ee bound}
\end{equation}
where we have quoted the more conservative limit.
This bound cannot be directly translated into a bound on $m_e$, since the relation between this and $m_{ee}$ is not unique, as it can be seen by comparing eqns. (\ref{eq:m_e}) and (\ref{eq:m_ee}), but instead depends on the values of the mixing matrix elements. This fact can be used to distinguish between different models. In particular, if $m_{ee}$ is found to be close to the upper bound (\ref{eq:m_ee bound}), as it has been recently claimed, the inverse and normal hierarchy scenarios would be strongly disfavored with respect to the degenerate one.

Finally, very recently the analysis of the cosmic microwave background anisotropy spectrum as observed by the Wilkinson Microwave Anisotropy Probe (WMAP), combined with the data on the galaxy power spectrum, has lead to the following bound on the sum of the three mass eigenvalues, roughly an order of magnitude more restrictive than the laboratory limit~\citep{Spe03}: 
\begin{equation}
\sum_{i=1}^3|m_i|<0.7\,\eV
\end{equation}
It should be noted however that, since many parameters contribute in giving to the cosmic background radiation and galaxy spectra their actual shape, this bound is model dependent, relying on several assumptions on the values of the other cosmological parameters, such as the total density $\Omega$. In effect, more conservative analyses lead to a somewhat higher limit, of the order of 1~eV. In the following we shall discuss in more detail the issue of the stability of the cosmological bound on neutrino masses when some assumptions, mainly the neutrino-antineutrino symmetry, are relaxed.
%As we have already noticed, the evidence for neutrino oscillations implies non vanishing neutrino masses; this in turn implies either the existence of rigth handed neutrinos, lepton number violation, or both.

\section{Cosmic Microwave Background Radiation}
The cosmic microwave background (hence thereafter simply CMB), namely the 2.7 K black body radiation that fills our Universe, was originated when baryons and photons decoupled and the latter began to propagate freely toward us. For this reason the CMB encodes a wealth of information about the Universe as it was at the epoch of decoupling, roughly 300.000 years after the Big Bang, corresponding to $z_\mathrm{dec}\sim1100$. 

The first striking feature of the CMB radiation (apart from it being a nearly perfect~\footnote{It is worth stressing the main limitation in detecting the deviation of the CMB spectrum from a black body one is our capacity of building a reference black body.} black body) is its extremely high degree of isotropy: the temperature varies by only a few parts in $10^5$ all over the sky. This means that the early Universe was indeed a very homogeneous and isotropic place, being well approximated by a Friedmann-Robertson-Walker (FRW) model.

However, much attention has to be paid to the small anisotropies in the CMB spectrum, since they are strictly related to the density perturbations at the epoch of decoupling that, growing, have originated the structures we observe today. The anisotropy spectrum can then be used to put very precise constraints on the overall properties of the Universe, i.e., to the cosmological parameters, since, as already noticed, the Universe at $z_\mathrm{dec}$ is a conceptually simple place, well described using the linear theory of perturbations on an homogeneous  and isotropic background space-time, and the physical processes relevant at that time (mainly the Compton scattering that keeps matter and radiation at equilibrium) are very well understood. In this section we describe how the anisotropies are generated; in the following we will explain how they can be used to estimate the cosmological parameters.

\subsection{Causes of temperature fluctuations}
The temperature fluctuations that we observe when looking at the CMB in different points in the sky, are usually classified according to their origin, in the following way. The \emph{primary anisotropies} are produced at the time of the decoupling; we will be mainly concerned with them. The \emph{secondary anisotropies} are produced by processes occurring in the way of the photons from the last scattering to us. One example is the integrated Sachs-Wolfe effect (ISW), occurring when the gravitational potential in which the photon moves is varying with time; this is the case in a $\Lambda$ dominated Universe. Finally, the \emph{tertiary anisotropies} are due to the fact that the CMB is not the only source of microwave radiation in the sky, so that its signal is contaminated by several foreground emissions. The most important are the emission from the galactic dust, the synchrotron emission from the electrons accelerated in magnetic fields, the thermal bremsstrahlung emission from hot electrons in the interstellar gas, and finally the emission from extragalactic point sources (AGN, blazars, quasars, ...). Although the tertiary anisotropies are not anisotropies intrinsic to the CMB, so that their contribution has to be subtracted when looking for the cosmological signal, nevertheless they contain valuable information on several astrophysical phenomena.
\subsection{Generation of primary anisotropies}
The primary anisotropies are produced by three basic mechanisms. The first one is directly related to the fluctuations in the matter density. Since before decoupling photons and baryons are tightly coupled, a fractional change in the density of the latter produces a  change in the density of the former, and this in turn produces a change in the temperature of the radiation: a denser region will look hotter, while a less dense one will look colder.

The second mechanism is the Doppler shift of the radiation. The peculiar velocity field of the baryons in the plasma is perturbed in a similar way to the density field; photons scattered by baryons that are moving toward us will look hotter, and vice versa.

The third mechanism is the Sachs-Wolfe (SW) effect. Photons coming from overdense region will be redshifted due to the gravitational redshift, and then look colder: those coming from underdense region will be blueshifted and look hotter.

Taking into account all these contributions, the equation for the fractional temperature fluctuation $\Delta T/T_0$ at a direction $\hat n$ is:
\begin{equation}
\frac{\Delta T}{T_0}=\frac14\,\delta_\gamma-\hat n\cdot\vec v+\frac13\,\phi
\end{equation}
where $T_0$ is the mean temperature, $\delta_\gamma$ is the fractional fluctuation in the photon energy density (the factor $1/4$ coming from the fact that for a black body $\rho_\gamma\propto T^4$), $\vec v$ is the velocity of baryons, $\phi$ is the gravitational potential, and all quantities are evaluated at the time of decoupling. It is the combination of these three terms that determines the main features of the field of temperature fluctuations; since they are model dependent, the pattern of anisotropies can be used to distinguish between different models.
\subsection{The Power Spectrum}
The quantity used to describe the statistics of temperature fluctuations is the \emph{power spectrum}. The temperature field in the sky is expanded in spherical harmonics:
\begin{equation}
\frac{\Delta T(\vec{n})}{T}=\sum_{l=1}^{\infty}\sum_{m=-l}^{l}a_{lm}Y_{lm}(\vec n)
\end{equation}
where
\begin{equation}
a_{lm}=\frac{1}{T_0}\int d\hat n\,T(\hat n)Y^*_{lm}(\hat n)
\end{equation}
are the multipole coefficients. The temperature angular power spectrum $C_l$ is then given by:
\begin{equation}
\langle a^*_{lm}a_{l'm'}\rangle=C_l\delta_{ll'}\delta_{mm'}
\end{equation}
Where the angle brackets $\langle\dots\rangle$ denote an average over a statistical ensemble. Unfortunately, our sky is just one realization of such ensemble; however, owing to the fact that the spectrum does not depend on $m$, we can construct an unbiased estimator of $C_l$ in this way:
\begin{equation}
C_l=\frac{1}{2l+1}\sum_{m=-l}^{l}a^*_{lm}a_{lm}
\end{equation}
A given cosmological model will predict the values of the $C_l$. This can then be used as a test for cosmological models.
%The spectrum of CMB anisotropies, thus, carries with it many informations on the distribution of matter at $z_\mathrm{dec}$, and, in turn, on the parameters that describe our Universe.

\section{Parameter estimation techniques}

The problem of parameter estimation is often stated stated as follows: given a set of experimental data, we want to find the set of parameters that provides the best fit to the data themselves. In other words, we want to find the set of parameters that has the maximum probability to produce the observed data. However this is just part of the story, since the probability that a given model produces the data can be put in relation with the inverse probability, namely the probability that the data are produced by a given model, only once we assign some \emph{a priori} probability do the different models (Bayes' Theorem). \\
Thus, defining a maximum likelihood function ${\cal L}({\mathbf d}|\,{\boldsymbol{\theta}})$, being equal, apart from a normalization constant, to the probability that the model defined by the vector of parameters~\footnote{In the following we shall simply say that $\boldsymbol{\theta}$ is ``the model''.}~$\boldsymbol{\theta}$ produces the data $\mathbf d$, Bayes' theorem states that:
\begin{equation}
\cal P(\boldsymbol \theta|\,\mathbf d)\propto{\cal L}({\mathbf d}|\,{\boldsymbol{\theta}})\cal \cdot P(\boldsymbol \theta) \label{Bayes}
\end{equation}
where $\cal P(\boldsymbol \theta|\,\mathbf d)$ is the probability that the data $\mathbf d$ come from the model $\boldsymbol{\theta}$, while $\cal P(\boldsymbol \theta)$ is the a priori probability, usually called \emph{prior}, assigned to the model $\boldsymbol{\theta}$.
If we are interested only to one parameter $\theta_i$ and not to the whole vector $\boldsymbol \theta$, we have to perform a marginalization of the distribution function over the remaining parameters:
\begin{equation}
{\cal P}(\theta_i|\,\mathbf d)\propto\int P(\boldsymbol\theta|\,\mathbf d)\, d\theta_1...d\theta_{i-1}d\theta_{i+1}...d\theta_n \label{marg}
\end{equation}
The maximum likelihood principle states that the best estimator $\hat{\boldsymbol\theta}$ of the set of parameters $\boldsymbol{\theta}$ is the value for which $\cal P(\boldsymbol \theta|\,\mathbf d)$ assumes its maximum value. The problem of parameter estimation is then reduced, at least in principle, to the knowledge of the function $\cal L$ and to the search of the maximum of the function defined in (\ref{Bayes}) with respect to the variable $\boldsymbol{\theta}$, being $\mathbf d$ known and fixed. In practice, however, such a search is not a simple task, in particular when one has to deal with a high dimensional parameter space. In the following we shall describe two techniques for the search of the maximum of the maximum likelihood function, supposing to be always able to compute ${\cal L}({\mathbf d}|\,{\boldsymbol{\theta}})$ for a given value of $\boldsymbol{\theta}$.

\subsection{`Brute force' approach}\label{subsec:brute}

In this approach, the parameter space is sampled using a grid (see \cite{Teg00} for an example of an application to the CMB data). Let $D$ be the number of dimensions in the parameter space (equal to the number of parameters) and let us consider, for the sake of simplicity, the same number $n$ of intervals along each dimension: the total number of points in the grid is then $n^D$. The function $\cal L$ is then computed at every point in the grid and is multiplied by the priors. Let us assume that the priors are uniform, so that we can `confuse' ${\cal L}({\mathbf d}|\,{\boldsymbol{\theta}})$ e $\cal P(\boldsymbol \theta|\,\mathbf d)$; the following is easy generalized to the case of non-uniform priors. Now, to obtain an estimation of the value of one of the parameters, we have to marginalize over the others, as said before. A direct integration is, however, computationally very expensive, and in many analyses it is substituted by a maximization over the remaining parameters. Let us consider for example a parameter $\omega=\theta_i$ ($i=1,...,D$); let $\omega_j$ be the values of $\omega$ sampled over the grid ($j=1,...,n$). The procedure is then the following (for simplicity purposes we omit the dependence from the data $\mathbf d$ in the argument of the distribution functions):
\begin{enumerate}
\item We fix the value of $\omega$ to a given $\bar\omega$;\\
\item We search for the maximum of ${\cal L}({\boldsymbol{\theta}})$ for $\omega=\bar\omega$ and assume that:
$$
\cal P(\omega=\bar\omega)\propto\max_{\omega=\bar\omega}\left\{ {\cal L}({\boldsymbol{\theta}})\right\} 
$$
\item We repeat the procedure for all the values $\omega_j$ ($j=1,...,n$) and thus obtain a $n$-point sampling of $\cal P(\omega)$;\\
\item We perform an interpolation to obtain a likelihood curve for $\omega$;\\
\item We repeat the procedure for the others parameters.
\end{enumerate}
Now we have a probability distribution function for every parameter, and the problem is reduced to the one of finding the maxima of $n$ functions of one variable, surely easier than its multi-dimensional analogous.

The theoretical reason for the approximation described at the point 2) lies in the fact that the result of the marginalization (\ref{marg}) and that of the maximization are exactly the same if the maximum likelihood function is a multivariate gaussian.\\
The main shortcoming of this approach is the fact that it requires large computational times. In the case of the CMB spectrum, the computation of the maximum likelihood function requires from 10 seconds to slightly more than a minute, on a typical workstation. If we consider a typical parameter space, with 7 dimensions, sampled in 10 point in each dimension, we obtain a grid made of $10^7$ points. Assuming we need 30 seconds to compute every model, the requested time is nearly 10 years. Using more powerful machines can reduce the computational times by some orders of magnitude, but the effort stays prohibitive for 10, or more, dimensional parameter spaces.

\subsection{Markov Chain Monte Carlo method}

The \emph{Markov Chain Monte Carlo} method (simply MCMC) makes easier the search for the maximum of the likelihood function, using a `clever' algorithm to sample the parameter space. It allows even, through the same algorithm, to compute with a lesser effort the marginalization integrals, avoiding in this way to approximate them with a maximization. Finally, it makes possible to explicitly compute the values of the normalization constants that we have overlooked in the preceding section. We shall briefly illustrate how this results are achieved; for further explanation see \cite{Chris01}.

The idea, upon which the MCMC method is based, is that of building up a Markov chain whose target distribution is the one we want to sample (in our case, the $\cal P(\boldsymbol \theta)$). This means that, once the Markov chain has converged, it becomes a random number generator distributed according to the $\cal P(\boldsymbol \theta)$. In this way it is possible to effectively sample the $\cal P(\boldsymbol \theta)$; the Markov chain will `visit' more often the points with higher probability, in a way proportional to the value of $\cal P$ in the point. The building of a Markov chain having as its target distribution the one we are interested to is made using a Metropolis-Hastings algorithm.

It is the clear that the main advantage of the MCMC method is that the research algorithm does not waste time in zones of low probability. This implies that the computational times scale with the number of parameters in a slower than exponential way.

\section{Neutrinos and Cosmology}
Our Universe is filled with a background of relic, non-interacting neutrinos, whose number density is of the order of 100 particles/cm$^3$ per species. Since, as pointed out by the recent experimental evidences, neutrinos are massive, they can give a non negligible contribution to the total energy density of the Universe. This contribution can be expressed, for a single species of mass $m$, as follows \cite{La03}:
\begin{equation}
\Omega_\nu h^2=\frac{m}{93\,\eV}A(\xi)
\label{eq:Omega_nu}
\end{equation}
where $h$ is the Hubble constant in units of 100~km~sec$^{-1}$~Mpc$^{-1}$.
%$\frac{\textrm{km}}{\textrm{sec Mpc}}$.
The function $A(\xi)$ is defined as:
\begin{equation}
A(\xi)=\frac{1}{4\eta(3)}\left[\frac13\,|\xi|^3+4\eta(2)|\xi|
+4\sum_{k=1}^\infty(-1)^{k+1}\,\displaystyle\frac{e^{-k|\xi|}}{k^3}\right]
\end{equation}
where $\eta(n)$ is the Riemann $\eta$ function of index $n$, and $\xi$ is a dimensionless parameter representing the chemical potential of the neutrino gas, normalized at the decoupling temperature. A non null chemical potential implies a lepton asymmetry, so we can think of it as quantity parameterizing such an asymmetry. In the case of no lepton asymmetry, equation (\ref{eq:Omega_nu}) reduces to the well known Gerstein-Zel'dovich formula.

The existence of a background of light ($m\ll1$~Mev) neutrinos acts in a different way on the observable cosmological quantities, that can then be used to extract informations on the properties of neutrinos \cite{Dol02}. Between others, one can get informations on the mass, on the lepton asymmetry, on the number of families, other than on the existence of sterile neutrinos, of radiative decays, and on the neutrino mean life.

In particular, the CMB spectrum is affected by the quantity $f_\nu\equiv\Omega_\nu/\Omega_m$, namely the ratio between the neutrino density and the total matter density. In effect, since neutrinos are ultra-relativistic at recombination, their presence shifts the time at which matter-radiation equality occurs, a parameter whose value affects the CMB spectrum. As we can see from eqn (\ref{eq:Omega_nu}), $f_\nu$ depends on the mass, the chemical potential and the number of families.

The matter power spectrum is itself affected by $f_\nu$, since the small-scale fluctuations are suppressed in a degree proportional to $f_\nu$, due to the high velocity dispersion (free streaming) of the neutrino gas.

Finally, the abundance of $^4\mathrm{He}$ produced during the primordial nucleosyntheys is affected both by the the total density of neutrinos, since this influences the expansion velocity and consequently the reactions freeze-out, and by the neutrino-antineutrino asymmetry in the electronic sector. Combining the nucleosynthesis data together with the CMB data, the following bounds are obtained \cite{Han02}:
\begin{eqnarray}
|\xi_{\mu,\tau}|<\,2.6\\
|\xi_{e}|<0.3
\end{eqnarray}
On the other hand, if the Large Mixing Angle solution for the mixing matrix is true, the oscillations equalize the chemical potentials and then \cite{Dol02b}:
\begin{equation}
|\xi_{\mu,\tau,e}|<\,0.07
\end{equation}
In this way, by obtaining and independent estimation of $\xi$, it would be in principle possible to falsify the LMA solution.

About the mass, the best estimation is the obtained using together the CMB data of the satellite \emph{Wilkinson Microwave Anisotropy Probe} (WMAP), of the \emph{Cosmic Background Imager} (CBI) e of the \emph{Arcminute Cosmology Bolometer Array Receiver} (ACBAR), the data on the large scale structure of the \emph{Two Degree Field Galaxy Redshift Survey} (2dFGRS) and the data on the matter power spectrum at $z\sim 3$ obtained by the observations of Ly$\alpha$ forest. The resulting upper bound, obtained in the hypothesis of three species with the same mass and null chemical potential, is (as already quoted in section \ref{sec:nuphys}) \citep{Spe03}:
\begin{equation}
m_\nu < 0.23\mathrm{eV}
\end{equation}
As we have already noticed, this represents a improvement of one order of magnitude with respect to the laboratory limits. However, some authors have criticized this result \citep{Elg03, Han03}, on the basis of the fact that the interpretation of the Ly$\alpha$ data is still controversial, and of the fact that there are some degeneracy effects in the paremeter space. In particular, in \citep{Elg03} the influence of the priors on $\Omega_m$ and on the bias parameter $b$ has been discussed , and it has been shown that the WMAP and 2dFGRS data are well fitted by a model having $\Omega_\nu=0.2$, $\Omega_m=1$ e $\Omega_\Lambda=0$, altough this would require a very low value for the Hubble constant, $h<0.5$. In a similar way, in \citep{Han03} the authors stress the degeneracy existing between the neutrino mass, the Hubble constant and the bias parameter. The latter is a phenomenological parameter that is introduced to take into account the fact that the galaxy distribution could not trace exactly the actual matter distribution, and then be biased. The main advantage in combining the CMB data with the data of the large scale structures, is that the former can be used to fix the amplitude of fluctuation, and then to put some bound on the bias. However, since as said before the effect of neutrinos is the suppression of the power on small-scales, the two effects can be in some limit confused.

Finally, the WMAP analysis assumes that the chemical potential of all species is equal to zero, and then that a perfect symmetry exists in all leptonic sectors. From the the theoretical point of view, there is no reason to suppose this.

\section{Effect of a lepton asymmetry on the parameter estimation}
Nearly all the analyses of the cosmic background spectrum assume, often implicitly, a perfect neutrino-antineutrino symmetry, and the a null chemical potential. Some exceptions exist, regarding the use of the CMB together with the abundance of $^4\mathrm{He}$ to put a bound on the chemical potential; however it is assumed that the neutrino has a null mass \citep{Han02, Man01}. In this way the effect of the chemical potential can be paremeterized in a simple way, introducing an effective number of neutrino families $N_{\nu,\,\mathrm{eff}}$. Furthermore, just two parameters are used: $\Omega_bh^2$ e $N_{\nu,\,\mathrm{eff}}$. In \citep{Les99}, the effect on the power spectrum of a non null neutrino mass and chemical potential has been studied, but the authors do not make a statistical analysis over the CMB data.

It is then worth studying the effect that a prior on $\xi$ can have on the determination of the other parameters, and trying to understand if there exists a degeneration between it and the other parameters. For this, we have performed a preliminar analysis of the BOOMERanG 98 (B98) data, using the techique described in the section \ref{subsec:brute}. In the following we shall describe it in more detail.

The first step is the choice of the parameter space. We have restricted our attention to flat models, with baryons, cold dark matter, cosmological constant and three neutrino species with the same mass and chemical potential. We have 6 parameters: the physical baryon density $\Omega_bh^2$, the density parameter associated to the cosmological constant $\Omega_\Lambda$, the scalar spectral index $n$, the Hubble constant $h$, the neutrino mass $m_\nu$, the neutrino chemical potential $\xi_\nu$. The values are listed in table \ref{tab: param}.
\begin{table}[h]
\begin{tabular}{lc}
\hline
\hline
\\
Parameter&Values\\[0.1cm]
\hline
\\
$\Omega_bh^2$&\{0.004, 0.009, 0.02, 0.045, 0.1\}\\[0.1cm]
$\Omega_\Lambda$ &\{0, 0.25, 0.5, 0.75, 1\}\\[0.1cm]
$h$&\{0.6, 0.7, 0.8, 0.9, 1.0\}\\[0.1cm]
$n$&\{0.90, 0.95, 1.0, 1.05, 1.10\}\\[0.1cm]
$m_\nu$& \{0.02, 0.033, 0.056, 0.093, 0.15, 
0.26, 0.43, 0.72, 1.20, 2\} eV\\[0.1cm]
$\xi_\nu$&\{0, 0.5, 1.0, 1.5, 2.0, 2.5,
 3.0, 3.5, 4.0, 4.5, 5.0\}\\
\hline
\end{tabular}
\caption{Parameter values}
\label{tab: param}
\end{table}

The second step requires to compute the power spectrum for every combination of the 6 parameters. The total number of possible combination is about 70,000, but some of them correspond to non-flat models. Eliminating these, we are left with about 50,000 models. The spectra are computed using the \verb+CMBFast+ code, modified according to \citep{Les99} to take the chemical potential into account.

The third step requires to process the theoretical spectra to obtain quantities comparable to the B98 data. For this we have to reduce the ${\cal C}_l$ to \emph{band power}, averaging over near, correlated multipoles through a window function. In this way we obtain uncorrelated points. The quantities obtained are called ${\cal C}_b$.

The fourth step is the computation of the likelihood of each model. We use the \emph{offset lognormal approximation} introduced in \citep{Bond00}:
\begin{equation}
-2\ln{\cal L}({\cal C}_b^{th})=\sum_{bb'}\frac{(\zeta_b^{th}-\zeta_b^{ex})}{\sigma_b}G_{bb'}\frac{(\zeta_{b'}^{th}-\zeta_{b'}^{ex})}{\sigma_{b'}}
\end{equation}
where
\begin{equation}
\zeta_b\equiv\ln({\cal C}_b+x_b)
\end{equation}
and $x_b,\,\sigma_b,\,G_{bb`}$ are known quantities.
Finally, the last step is the marginalization, using the maximization method described in section \ref{subsec:brute}. First we have considered only models having $\xi=0$, imposing then a prior on the chemical potentail; then we have considered the whole space. In this manner we obtain two likelihood curves for each parameter; we report them in in figs. 1 and 2. In fig. 1, we have just plotted the sampled points, and joined them by lines. In fig. 2 we show a tentative interpolation.

\begin{figure}[t]
\begin{center}
\includegraphics{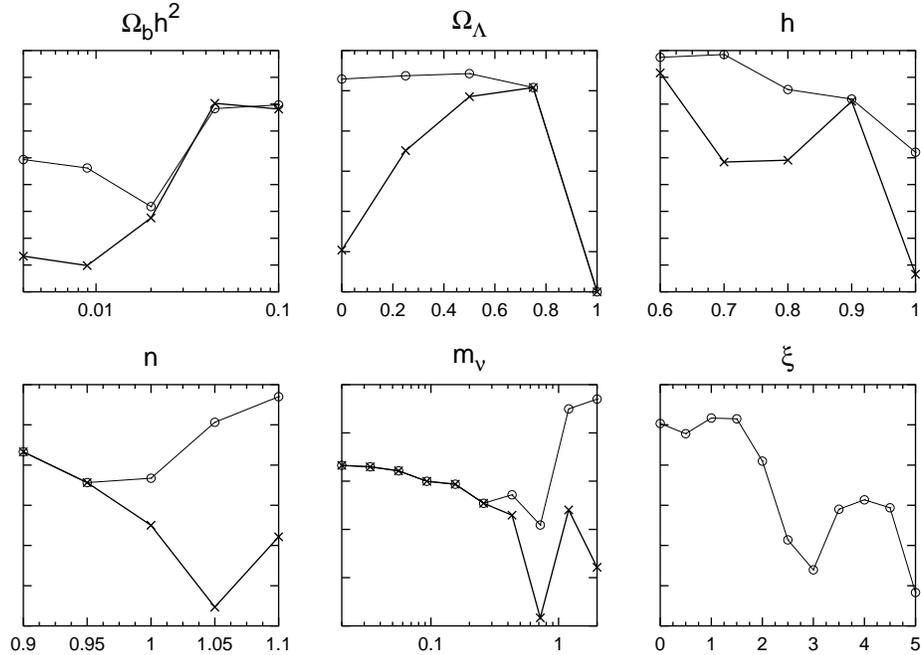}
\caption{Sampling of the likelihood curves for the six parameters we have considered. Upper row, from left to right: the physical baryon density $\omega_b$, the vacuum energy density $\Omega_\Lambda$, the Hubble constant $h$. Lower row, from left to right: the scalar spectral index $n$, the neutrino mass $m_\nu$, the chemical potential $\xi$. The thick lines correspond to the case with $\xi=0$, the thin lines correspond to the absence of assumptions on $\xi$. }
\end{center}
\end{figure}

\begin{figure}[t]
\begin{center}
\includegraphics{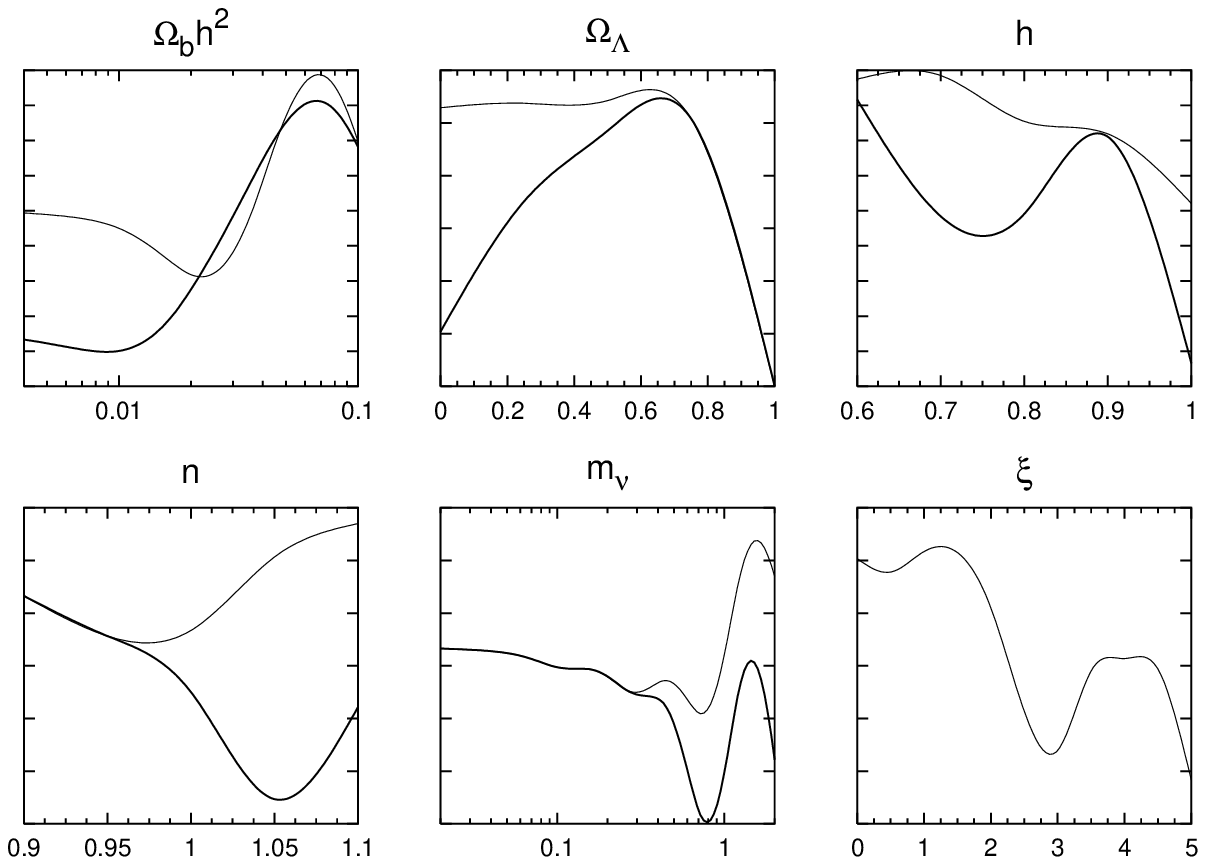}
\caption{The same as in fig. 1, but with a tentative interpolation.}
\end{center}
\end{figure}
 
The comparison between the two curves provides useful information on the effect of lepton asymmetry on the cosmological parameter estimation. We expect to have, in the case $\xi=0$, an agreement with WMAP's best fit values (but please note that we did not include the optical depth $\tau$ in our parameters space). Now let us discuss separately each parameter.

The physical baryon density $\omega_b\equiv\Omega_bh^2$ shows a maximum at about 0.03--0.04, both for $\xi=0$ and for freely vaying $\xi$. In this last case, however, the low baryon density models are less disfavoured than in the former case. WMAP's best fit value is 0.024.

The vacuum energy density parameter $\Omega_\Lambda$ shows, in the case $\xi=0$, a clear peak at about 0.7. When we allow $\xi$ to vary, the curve get flat between 0 and 0.7. Such a behaviour seems to point to a degeneracy between $\xi$ e $\Omega_\Lambda$. In \citep{Lu02}, a similar effect has been observed, with respect to the data on the large scale structures. It seems then worth looking deeper into this problem. WMAP's best-fit value is 0.7.

The dimensionless Hubble constant $h$ shows, when $\xi=0$, a maximum between 0.6 and 0.8. When $\xi$ can vary, the maximum is shifted toward the left. WMAP's best-fit value is 0.72.

For the scalar spectral index $n$, low ($n<0.95$) values are favoured when $\xi=0$; when $\xi$ is freely varying, higher ($n>1.05$) values seem to be preferred. WMAP's best fit value is 0.99.

For the neutrino mass $m_\nu$, low ($m_\nu<0.5$~eV) values are favoured when $\xi=0$; when $\xi$ is allowed to vary, a peak appears at 1~eV. The WMAP team put an upper bound to the neutrino mass of $m_\nu<0.23$~eV.

The dimensionless chemical potential $\xi$ shows two maxima: the first at 1.5, the second at 4. We note that the latter is not in agreement with the upper limit of 2.6 imposed by the primordial nucleosynthesis.

\begin{figure}[t]
\begin{center}
\includegraphics[width=1.2\hsize,clip]{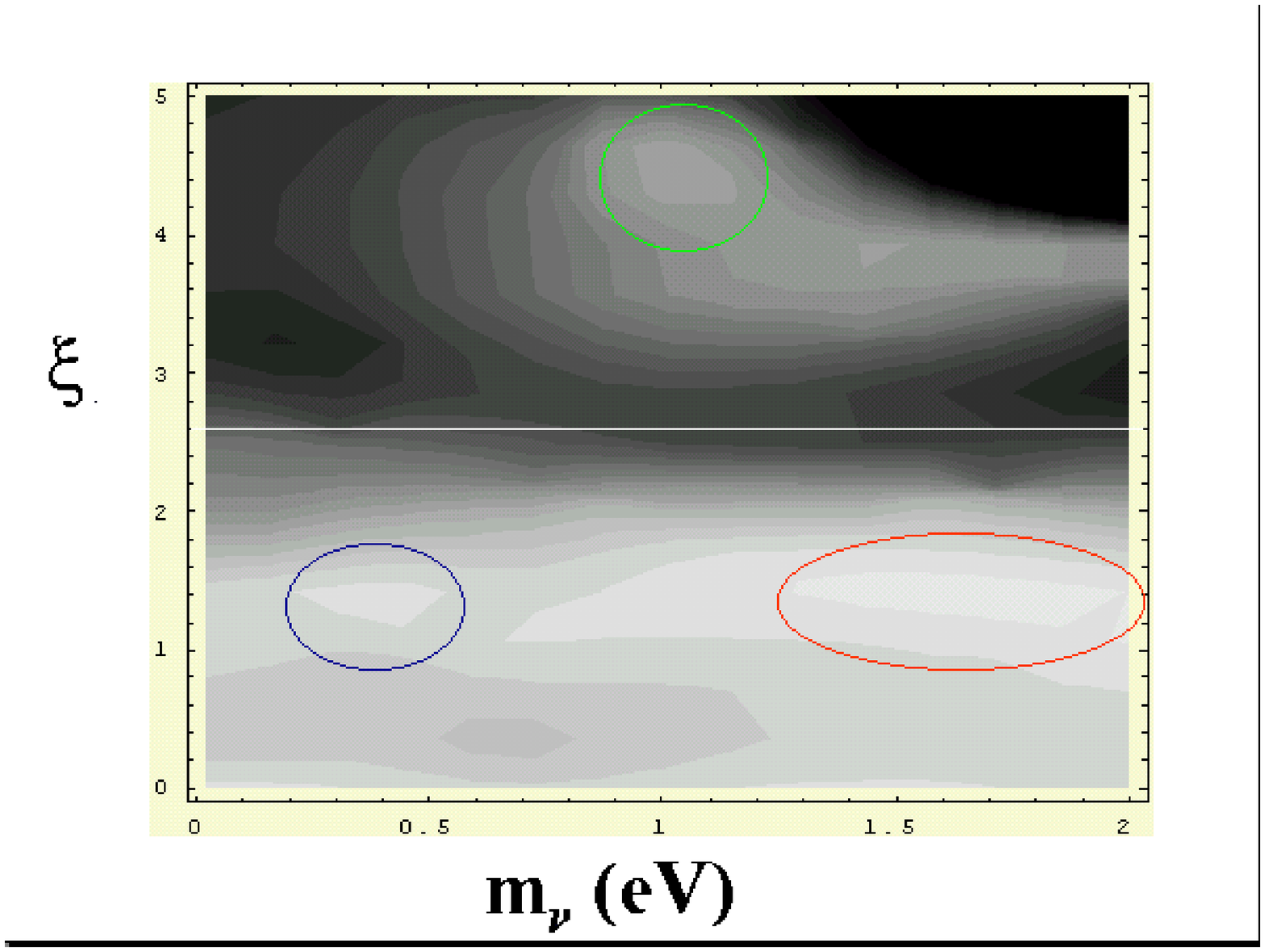}
\end{center}
\caption{Likelihood surface for $m_\nu$ e $\xi$. Lighter zones correspond to an higher likelihood. We have circled the local maxima. The white line is the upper limit $\xi<0.26$ imposed by the nucleosynthesis.}
\label{fig: mnuxi}
\end{figure}

Finally, in fig \ref{fig: mnuxi} we show the maximum likelihood surface for the pair $\{m_\nu,\,\xi_\nu\}$. We see three maxima, one at \{0.5 eV, 1.5\}, one at \{1.5 eV, 1.5\} and one at \{1 ev, 5\}. This last one would be excluded on the basis of the upper bound cited above.

%Le stime che sembrano maggiormente alterate sono quella sulla costante cosmologica, di cui gi\`a si \`e detto, quella sulla massa del neutrino e quella sull'indice spettrale. In particolare, quella sulla massa del neutrino sembra contraddire il limite posto dal team di WMAP.

\section{Conclusions}
The analysis described in the preceding section points to the fact that the assumptions usually made on the lepton asymmetry, do really influence the estimation of cosmological parameters. In particular, the parameters that are more affected are the cosmological constant, the neutrino mass and the spectral index. Values of the cosmological constant between 0 and 0.7 cannot be distinguished, having nearly the same probability. The neutrino mass is peaked at 1~eV, this being not in agreement with the WMAP results. Finally, higher values of the spectral index are favoured. We think then that is worth studying in more depth this topic. It is without doubt necessary to apply our method to the WMAP data, using a larger parameter space, to be explored using MCMC methods. 

\section*{Acknowledgements}
We wish to thank Remo Ruffini for his valuable advice and for having suggested the topic of this paper.

%\bibliography{biblio}

\end{document}